# Multifrequency and multimode topological waveguides in a Stampfli-triangle photonic crystal with large valley Chern numbers

*Bei Yan, Yiwei Peng, Jianlan Xie, Yuchen Peng, Aoqian Shi, Hang Li, Feng Gao, Peng Peng, Jiapei Jiang, Jianjun Liu*, Fei Gao*, and Shuangchun Wen*

B. Yan, Y. Peng, J. Xie, Y. Peng, A. Shi, H. Li, F. Gao, P. Peng, J. Jiang, J. Liu, S. Wen
Key Laboratory for Micro/Nano Optoelectronic Devices of Ministry of Education & Hunan Provincial Key Laboratory of Low-Dimensional Structural Physics and Devices, School of Physics and Electronics, Hunan University, Changsha 410082, China
E-mail: jianjun.liu@hnu.edu.cn (Jianjun Liu)

Y. Peng, F. Gao
State Key Laboratory of Modern Optical Instrumentation and The Electromagnetics Academy at Zhejiang University, Zhejiang University, Hangzhou 310027, China
E-mail: gaofeizju@zju.edu.cn (Fei Gao)

B. Yan
Department of Electrical and Electronic Engineering, Southern University of Science and Technology; Shenzhen 518055, China

The multifrequency quantum valley Hall effect (QVHE) with a large valley Chern number has been realized to significantly improve the transmission capacity of topological waveguides and increase the mode density of topological waveguides. However, multifrequency and multimode QVHEs have not been realized simultaneously. In this work, using tight-binding model calculations and numerical simulations, a valley photonic crystal (VPC) consisting of a Stampfli-triangle photonic crystal is constructed, and its multiple degeneracies in the low-frequency and high-frequency bands split simultaneously to realize the QVHE with multiple topological edge states (TESs). The multifrequency and multimode topological transmission with two low-frequency modes and four high-frequency modes is realized by means of simulations and experiments through a Z-shaped waveguide constructed using two VPCs with opposite valley Chern numbers to prove the realization of a large valley Chern number in the two frequency bands. The two low-frequency modes are successfully distinguished with position-dependent selective excitations, which experimentally demonstrates the occurrence of a large valley Chern number. A frequency-dependent multimode beam splitter is theoretically proposed for high-performance integrated photonic device applications. These results provide new ideas for high-efficiency and high-capacity optical transmission and communication devices and their integration; furthermore, they broaden the application range of TESs.

## 1. Introduction

In analogy to topological insulators in condensed-matter systems, photonic topological insulators and their micro–nano photonic devices have been proposed,[1–4] such as optical waveguides,[5] lasers,[6] beam splitters,[7] and logic gate optical circuits,[8] owing to their robust, low-loss topological properties. Each topological phase possesses its own topological invariant, [9,10] such as the Chern number to characterize the quantum Hall effect (QHE),[5,10–12] the $Z_2$ topological invariant (or spin Chern number) to characterize the quantum spin Hall effect (QSHE),[13–16] and the valley Chern number to characterize the quantum valley Hall effect (QVHE).[2,6,7,17–26] According to the bulk-edge correspondence principle,[9] the number of topological edge states (TESs) and the modes at the waveguide interface are equal to the difference between the topological invariants of the two structures constituting the waveguide. Skirlo *et al*. proposed that the Chern number could be an integer greater than 1; they achieved a QHE with a large Chern number and its multimode topological transmission by breaking multiple degeneracies.[11,12] However, at present, the realization of the QHE generally uses gyromagnetic materials and breaks the time reversal symmetry (TRS) through an external magnetic field, which is only applicable to the microwave frequency band, thus limiting its practical application in the optical frequency band. A recent study achieved a Chern number of 4 in a dielectric; this study involved gyroelectric materials and a complex three-dimensional (3D) structure.[27]

In order to overcome this limitation, all-dielectric valley photonic crystals (VPCs) [6,7,18–26,28] and surface plasmon crystals [29,30] have been proposed to realize the QVHE; these crystals were mainly based on the triangular lattice,[6,7,18,19,24,29,30] honeycomb lattice,[20,21,23,25,28] and kagome lattice.[22] In the abovementioned studies, the $C_6$ symmetry of the structure was broken, while the $C_3$ symmetry was preserved by changing the shape and size of the cylinders and expanding or shrinking the lattice, respectively, to break the degeneracy at the highly symmetrical point K (K′). Consequently, the non-zero and opposite Berry curvature distributions were obtained in the two regions around the K and K′ points in the Brillouin zone (BZ), and the valley Chern numbers were introduced. In 2020, Xi *et al.* realized a broken sublattice by expanding or shrinking the sublattices of the honeycomb lattice, breaking multiple degeneracies at non-high-symmetry points, to realize multimode topological transmission with a tunable, large valley Chern number.[23] Bilayer surface plasmon crystals have been proposed to obtain a Chern number of 2.[30] In 2021, Lu *et al.* reported a valley Chern number of 3 and a total Chern number of 4 in twisted multilayer graphene.[31] However, these designs worked only in the case of a single frequency band, which limits their potential applications in multifrequency waveguides, filters, and communication fields. There is also a multifrequency topological transmission and dual-band topology with a QVHE in the high-frequency band and QHE in the low-frequency band observed in gyromagnetic photonic crystals.[32] Although the multifrequency topological transmission has been studied based on VPCs,[20,24,25] it is a single-mode transmission. Therefore, in order to improve the mode density of topological waveguides as well as successfully realize high-efficiency and high-capacity multichannel integrated photonic devices, systematic investigations of the multifrequency and multimode topological transmission with a large valley Chern number are urgently required. Furthermore, two-dimensional (2D) photonic quasicrystals (PQCs) have a high symmetry and rich band structures,[33] and it has been certified that they can generate TESs,[16,34] which reveal that PQCs have great potential to simultaneously achieve a multifrequency and multimode topological transmission.

In this work, the basic structural unit of a Stampfli-type PQC was used as the unit cell in a triangular lattice arrangement to construct a Stampfli-triangle photonic crystal (STPC). The degeneracies in two frequency bands split simultaneously by changing the size of the cylinders in the unit cell of the STPC, which aids the realization of the QVHE with a number of TESs greater than 1 in the two frequency bands. Consequently, the QVHE with a large valley Chern number was realized in two frequency bands. The numerical simulation results are in good agreement with the theoretical calculations. Furthermore, the realization of the multifrequency and multimode topological transmission was confirmed through simulations and experiments by constructing a Z-shaped waveguide. This study provides a theoretical basis for obtaining multifrequency and multimode, high-performance integrated photonic devices. Finally, a frequency-dependent multimode beam splitter is theoretically proposed.

## 2. Results and Discussion

The STPC structure proposed in this work is shown in **Figure 1**.

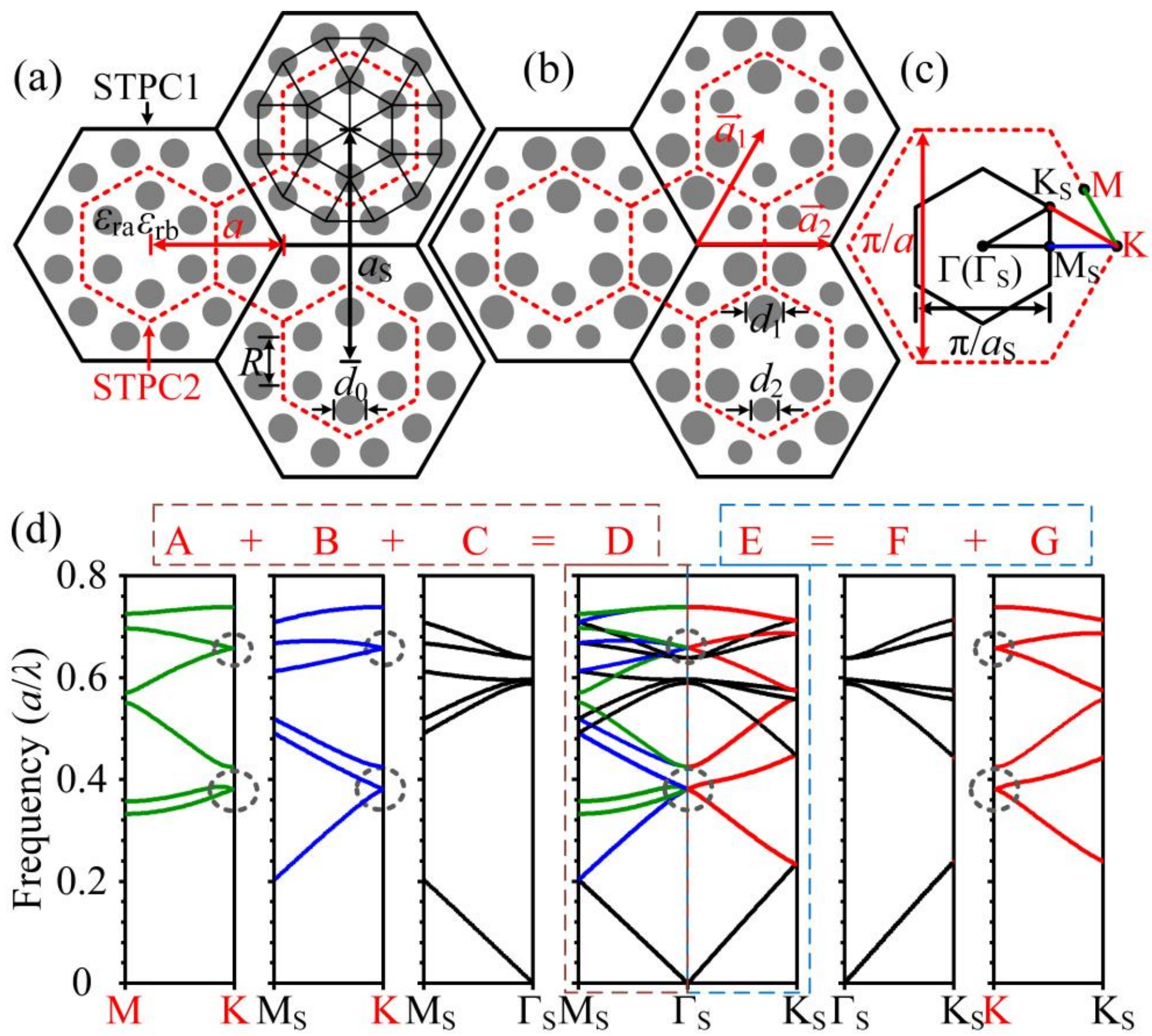

**Figure 1.** a) STPC structure with $d_0 = 0.6 \times R$. b) STPC structure with $d_1 = d_0 + \delta$ and $d_2 = d_0 - \delta$, where $d_0 = 0.6 \times R$ ($\delta = 0.1 \times R$). c) First BZs of STPC1 and STPC2 and distribution of the high-symmetry points. d) Band structures of STPC1 and STPC2.

The basic structural unit of the Stampfli-type PQC is arranged in a triangular lattice to construct the STPC; its lattice constant is $a_S = \sqrt{3}a$. As shown in Figure 1a, after removing the central cylinder in the unit cell of the STPC (STPC is denoted as STPC1 at this time), its structure can be determined by considering the unit cell consisting of the six cylinders in the dotted red frame arranged in a triangular lattice (STPC is denoted as STPC2 at this time), and the lattice constant of STPC2 is $a = 32.8$ mm (this macroscale is considered for the convenience of the experiment). The distance between the adjacent cylinders is $R = a_S/(3 + \sqrt{3}) = a/(1 + \sqrt{3}) = 12$ mm; the alumina cylinders (relative dielectric constant $\varepsilon_{ra} = 8$) with initial diameter $d_0$ are placed in air ($\varepsilon_{rb} = 1$). The diameter of the cylinders is adjusted and divided into two sets of diameters, namely $d_1 = d_0 + \delta$ and $d_2 = d_0 - \delta$, as shown in Figure 1b. A VPC can be constructed with a specific size and material to break the parity reversal symmetry (PRS). The solid black and dotted red frames in Figure 1c are the first BZs of STPC1 and STPC2, respectively. The band structures of STPC1 and STPC2 for the transverse magnetic (TM) polarization are shown in Figure 1d. The band structure D (E) calculated for STPC1 can be obtained by adding the band structures A, B, and C (F and G) calculated for STPC2.[34] The quadruple degeneracy at point $\Gamma_S$ of the band structure of STPC1 can be represented as two double degeneracies at the point K of the band structure of STPC2. The double degeneracy at the K point is more suitable for constructing the valley system.[22] Therefore, the VPC constructed by regulating the cylinder size of STPC2 can be used to replace STPC1 for further research.

The calculation of the valley Chern number was initiated from the Chern number of the QHE. The Chern number is associated with the Berry curvature $F_n(k)$, and the Chern number in each band can be expressed [9] as follows:

$$C_n = \frac{1}{2\pi}\int_{BZ} F_n(k) d^2k \, . \tag{1}$$

For a VPC that breaks the PRS and maintains the TRS, its Chern number $C_n = 0$. Non-zero and oppositely distributed Berry curvatures appear in the K and K′ regions. Additionally, when the degeneracies at the K and K′ regions split, each split degeneracy contributes a Berry flux of π, and the K and K′ regions acquire an opposite Berry flux. The valley Chern numbers $C_V = C_K - C_{K'}$, and the $C_K$ and $C_{K'}$, respectively, can be obtained by changing the integral region in Equation (1) to the half Brillouin zone corresponding to the K and K′ regions, respectively.[17] Previous studies [21,22] adopted the method reported in Reference [35–37], which discretized the Brillouin zone, and the valley Chern number was calculated based on Equation (1). However, their results showed that the valley Chern number is not an integer but varies with a change in the structural parameters. The valley Chern number is not an integer, nor it is equal to the number of TESs or topological waveguide modes. Therefore, it is difficult to use the valley Chern number as the most important criterion for judging multimode topological transmission; nonetheless, it can be through the distribution of the Berry curvature that it can be determined whether the TES and the topological transmission can be generated.

In order to realize the QVHE with a large valley Chern number, we calculated the band structures of STPC2 and the VPC, as shown in **Figure 2**. We also predicted the band structures based on the tight-binding Hamiltonian, as shown in Section I of the Supporting Information.

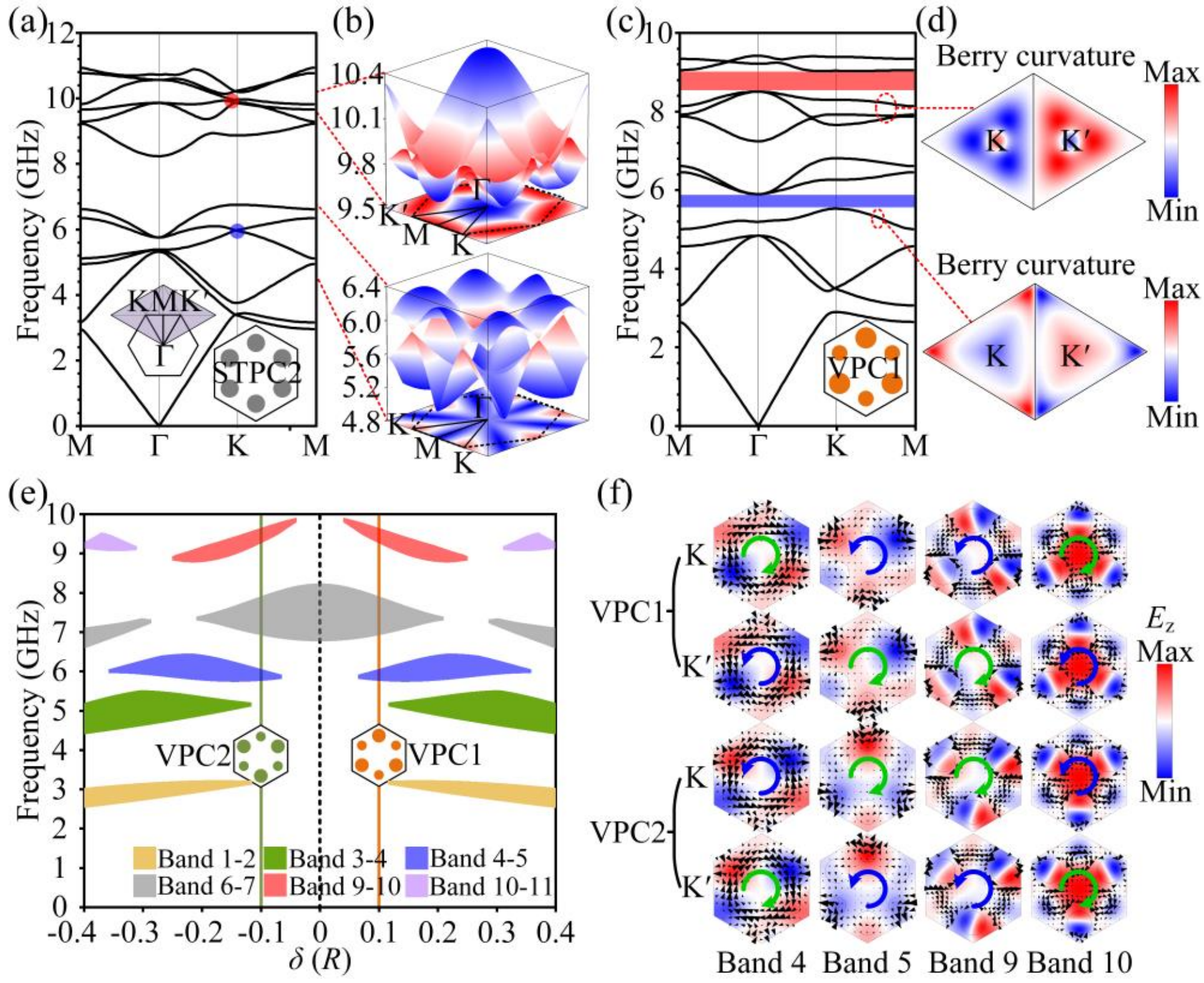

**Figure 2.** a) Band structure of STPC2; b) 3D band structures; c) Band structure of VPC1; d) Berry curvature distributions of VPC1 corresponding to the fourth band and the seventh to ninth bands; e) Bandgap distribution for different $\delta$ values at $d_0 = 0.6 \times R$; f) Electric field distributions and Poynting energy flows (black arrow) at points K and K′ of VPC1 ($\delta = 0.1 \times R$) and VPC2 ($\delta = -0.1 \times R$). When the Poynting energy flow direction is clockwise (anticlockwise), as marked by the green (blue) circular arrows, the vortex chirality corresponds to a valley pseudospin down (up).

As shown in Figure 2a-b, when $d_0 = 0.6 \times R$ and $\delta = 0$ for STPC2, there are degeneracies at the high-symmetry point K of the fourth and fifth bands at the non-high-symmetry point (the point between Γ and K) of the ninth and tenth bands. By tuning $\delta$ from 0 to $0.1 \times R$ (VPC1), the degeneracies between the fourth and fifth bands as well as the ninth and tenth bands are broken, as shown in Figure 2c. There are non-zero and opposite Berry curvature distributions for the K and K′ valleys, as shown in Figure 2d. It can be seen that VPC1 has non-zero valley Chern numbers in both the low-frequency and high-frequency bands. The band structures of VPC2 and VPC1 are the same, as shown in Figure 2e, but their valley Chern numbers and topological properties are opposite, as can be determined from the electric field distribution and the opposite Poynting energy flow directions of the VPC1 and VPC2 indicated in Figure 2f. The opposite pseudospin properties of the K and K′ valleys can be determined from the electric field distribution of VPC2 excited by the chiral source in the real and momentum spaces is shown in Section II of the Supporting Information.

The projected band structures were calculated based on the supercell constructed with VPC1 and VPC2 as well as the corresponding TESs and the electric field distributions at specific frequencies, as shown in **Figure 3**.

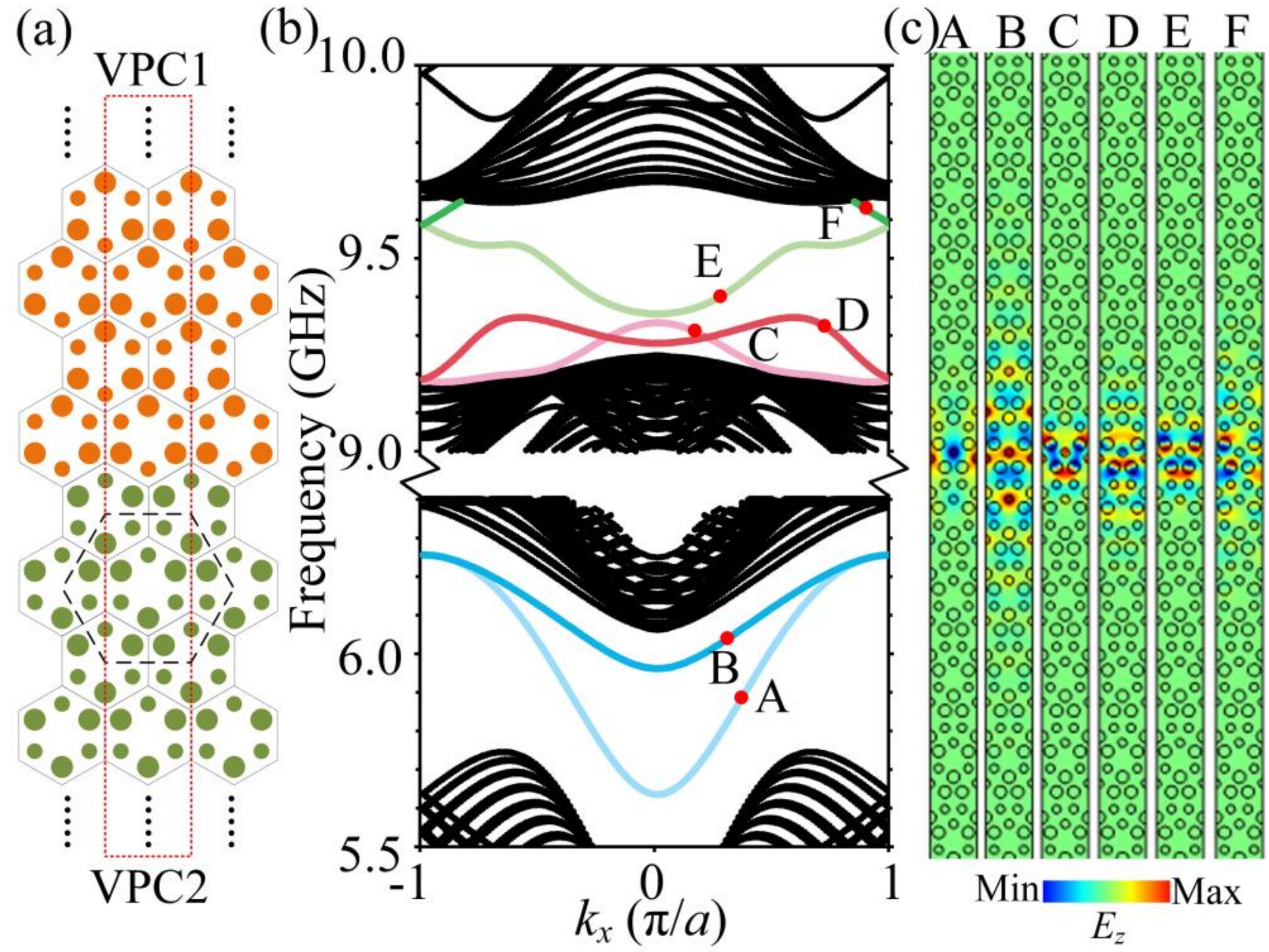

**Figure 3.** a) Supercell; b) Projected band structure; c) Electric field distribution at points A, B, C, D, E, and F.

Figure 3a shows that there are 15 unit cells of VPC1 and VPC2 each (not all shown), which form a vertically elongated supercell in the dashed red frame. Figure 3b illustrates that the numbers of TESs in the low-frequency and high-frequency bands are 2 and 4, respectively.

The number of TESs in the interface between the two VPCs will depend on the bulk topology for the VPCs and the geometry of the interface. We have calculated projected band structures based on the supercell constructed with VPC1 and VPC2 with different geometry of the interface, and the invariance number of TESs indicated that it is solely determined by the valley Chern number (more details see Section III of the Supporting Information). Clearly, the difference between the valley Chern numbers of the K (K′) valley in the low-frequency (high-frequency) band for VPC1 and VPC2 is represented as follows: $\left|\Delta C_{\mathrm{K(K')}}^{\mathrm{Low}}\right| = \left|C_{\mathrm{K(K')}}^{\mathrm{VPC1}} - C_{\mathrm{K(K')}}^{\mathrm{VPC2}}\right| = 2$ ($\left|\Delta C_{\mathrm{K(K')}}^{\mathrm{High}}\right| = \left|C_{\mathrm{K(K')}}^{\mathrm{VPC1}} - C_{\mathrm{K(K')}}^{\mathrm{VPC2}}\right| = 4$). The valley Chern numbers of VPC1 and VPC2 are equal in size and opposite in sign; thus, the valley Chern number in the low-frequency (high-frequency) band is given by $C_{\mathrm{V}}^{\mathrm{Low}} = C_{\mathrm{K}}^{\mathrm{Low}} - C_{\mathrm{K'}}^{\mathrm{Low}} = \pm 2$ ($C_{\mathrm{V}}^{\mathrm{High}} = C_{\mathrm{K}}^{\mathrm{High}} - C_{\mathrm{K'}}^{\mathrm{High}} = \pm 4$). The TESs at the interface of the supercell structure could not connect the upper and lower bands of the bandgap; [29,39] that is, the TESs could not be gapless. Furthermore, in analogy to the QSHE based on the $C_6$ symmetry, the $C_6$ symmetry broken at the interface leads to a gap in TES,[16] and the $C_3$ symmetry broken at the interface between VPC1 and VPC2 leads to gapped TESs.[23] Figure 3c demonstrates different electric field distributions at points A and B (C, D, E, and F) indicating that there are two (four) different topological edge modes. In other words, the number of TESs is equal to the number of topological waveguide modes.

Furthermore, herein, the multifrequency and multimode features were demonstrated through experiments and simulations. VPCs consisting of alumina cylinders (same structural and material parameters as in the simulation, height $h$ = 10 mm) were placed between two flat metallic plates (**Figure 4**a–c) to mimic the 2D theoretical model. Figures 4d–i illustrate that the stable anti-bending transmission in the low-frequency and high-frequency bands verifies the realization of the multifrequency and multimode topological transmission. Bulk and waveguide modes were excited by an L-shaped dipole antenna (inset of Figure 4a) connected to a vector network analyzer (Rohde & Schwarz ZVL13), and they were measured using another L-shaped probe.

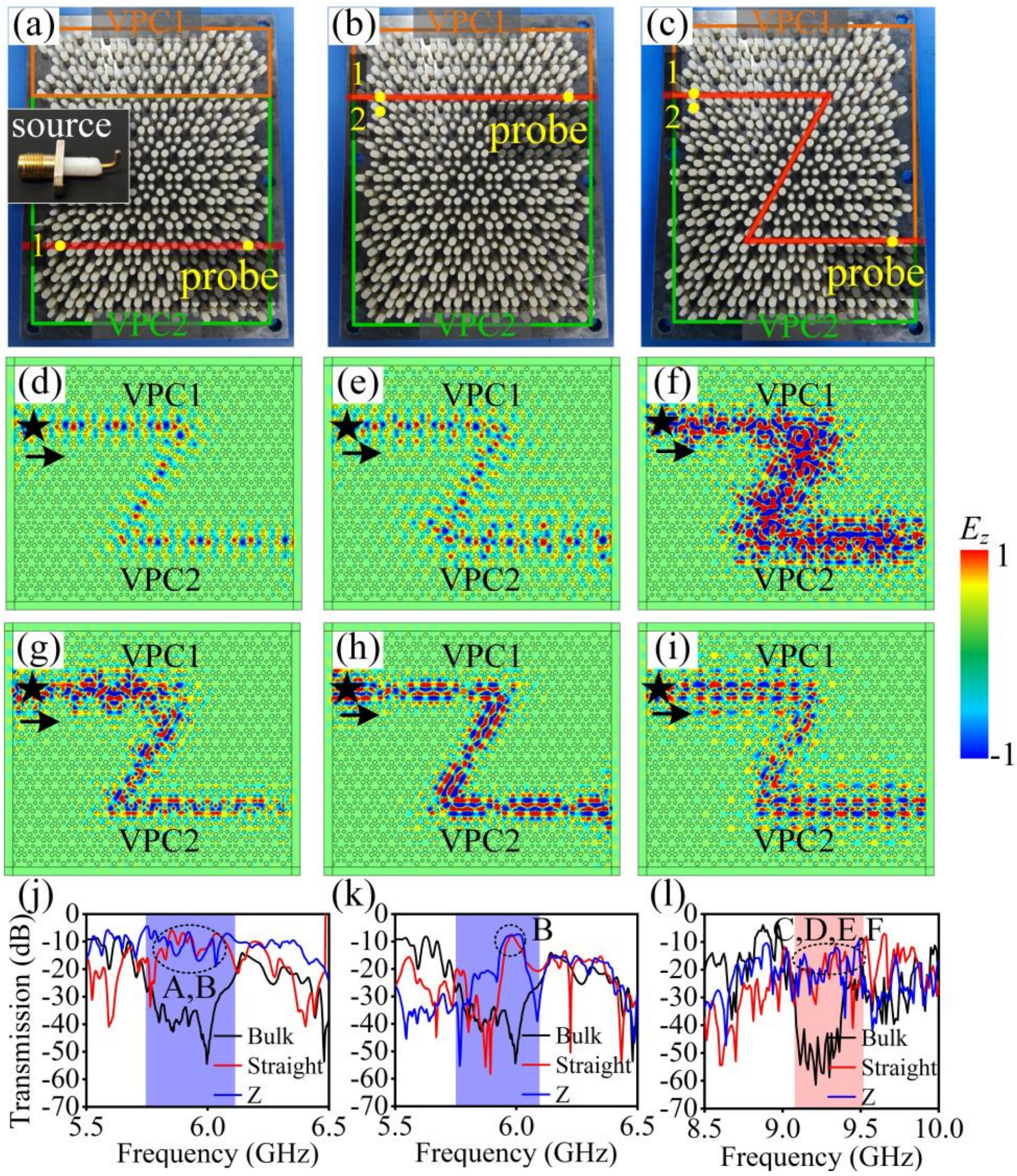

**Figure 4.** Experimental setups: a) bulk; b) straight waveguide; and c) Z-shaped waveguide. The inset shows the L-shaped dipole antenna, and the yellow dots indicate the excitation source and probe positions. Simulation of the Z-shaped waveguide propagation at different frequencies at the points A–F shown in Figure 3c (where the black star represents the point source $E_Z$): d) 5.9 GHz; e) 6.04 GHz; f) 9.29 GHz; g) 9.33 GHz; h) 9.4 GHz; i) 9.62 GHz. j) Measured transmission spectra of the low-frequency band with the source at position 1 in Figure 4a–c. k) Measured transmission spectra of the low-frequency band with the source at position 2 in Figure 4a–c. l) Transmission spectrum of the high-frequency band with the source at position 1 in Figure 4a–c.

Following the above mentioned setups, herein, the source antenna was first placed at position 1, and then the transmission spectra were measured. For both bands, the transmissions of the Z-shaped waveguide were found to be similar to those of the straight one and significantly higher than those of the bulk (Figure 4j–l). These measurements verify the multifrequency topological transmissions against sharp corners. More specifically, each transmission band is the contribution of multiple topological edge modes. For the lower-frequency band, the source at position 1 excites both modes A and B, due to the overlapping of their fields when the source is at position 1. However, modes A and B are markedly distinct in their transverse evanescent length, as shown in Figure 3c, 4d, and 4e. Furthermore, the source was placed at position 2, and the transmission spectra were measured, as shown in Figure 4k. Only the field of mode B is distributed, which indicates that mode A cannot be excited at position 2, where its eigenstate does not exist. In this case, the simulated dispersion bands shown in Figure 3b indicate that mode A occupies a wider frequency range than mode B, which is consistent with the measured narrower transmission band of mode B, as shown in Figure 4k. These measurements verify the multimode topological transmissions of the low-frequency band. Regarding the high-frequency band, the simulated eigenstates (Figure 3c) and the excitation field (Figure 4f–i) C, D, E, and F exhibit a similar field distribution; that is, they cannot be distinguished. Therefore, in the experiment, the position-dependent selective excitations are not applicable; thus, the source is only set at position 1, and the four modes from the measured transmission spectrum cannot be distinguished, as shown in Figure 4l. However, there are bandgaps for C, D, E, and F modes, every mode will not occupy all bandgap for bulk, and the measured transmission spectrum shows no bandgap for the edge state indicating that there is not only one mode. This illustrates that the simulation and experimental results are consistent with each other. Therefore, the multifrequency and multimode topological transmission can also be verified.

Although the modes at high frequencies are difficult to distinguish, the proposed VPC is essentially topologically different from previous designs of broadband TESs.[40–42] Benefiting from the two advantages of the proposed VPC, i.e., the large valley Chern number and the different topological properties in the two bandgaps, we can design a frequency-dependent multimode beam splitter (see Section Ⅳ of the Supporting Information), which has two working regimes: firstly, at high frequencies, it can realize beam splitting at the same frequency; secondly, when high-frequency and low-frequency beams exist simultaneously, beam splitting at multi-frequencies can be realized. It is difficult for VPCs with a small valley Chern number and a single bandgap [40] to exhibit both functionalities in one structure. Such a device is expected to be used in high-performance beam splitter applications as a primary element.

## 3. Conclusion

In this work, the simplest cell of the STPC structure was obtained, the PRS was broken by adjusting the diameter of the cylinders, and the broken degeneracies for the high-symmetry and non-high-symmetry points in different frequency bands were realized. The non-zero and anti-symmetric Berry curvature distributions were obtained for the K and K′ valleys. Two and four TESs were found in the low-frequency and high-frequency bands by constructing a rectangular supercell and calculating its projected band structure. Finally, a Z-shaped waveguide was constructed using simulations and experiments to achieve a 120° bending transmission, which proves the existence of TESs. Therefore, the possibility of realizing a multifrequency and multimode topological transmission with a large valley Chern number was verified. To demonstrate multifrequency, multimode, high-performance integrated photonic device applications, a frequency-dependent multimode beam splitter was theoretically proposed.

**Supporting Information**

Supporting Information is available from the Wiley Online Library or from the author.

**Acknowledgements**

This work was supported by the National Natural Science Foundation of China (Grant Nos. 61405058, 62075059, and 62171406), the Natural Science Foundation of Hunan Province (Grant Nos. 2017JJ2048 and 2020JJ4161), the Key project of Scientific Research of Hunan Provincial Education Department (Grant No. 21A0013), and the Fundamental Research Funds for the Central Universities (Grant No. 531118040112). The authors greatly acknowledge Professor Jian-Qiang Liu for software sponsorship and Professor Wei E. I. Sha, Yuanzhen Li from Zhejiang University, and Dr. Linyun Yang from Southern University of Science and Technology for very helpful discussions. Bei Yan and Yi-Wei Peng contributed equally to this work.